\documentclass{emulateapj}

\def\amin{\ifmmode^{\prime}\else$^{\prime}$\fi}
\def\asec{\ifmmode^{\prime\prime}\else$^{\prime\prime}$\fi}

\shorttitle{Nuclear Accretion in M32}
\shortauthors{Seth}

\begin{document}

\slugcomment{Accepted by {\em ApJ}, Oct. 2nd, 2010}

\title{Gas Accretion in the M32 Nucleus: Past \& Present}

\author{Anil C. Seth\altaffilmark{1}}

\altaffiltext{1}{Harvard-Smithsonian Center for Astrophysics, 60 Garden Street Cambridge, MA 02138, {\em OIR Postdoctoral Fellow}}

\begin{abstract}

Using adaptive optics assisted Gemini/NIFS data, I study the present
and past gas accretion in the central 3\asec~of the M32 nucleus.
From changes in the spectral slope and CO line depths near the center,
I find evidence for unresolved dust emission resulting from BH
accretion.  With a luminosity of $\sim2\times10^{38}$~erg$\,$s$^{-1}$,
this dust emission appears to be the most luminous tracer of current
BH accretion, two orders of magnitude more luminous than previously
detected X-ray emission.  These observations suggest that using high
resolution infrared data to search for dust emission may be an
effective way to detect other nearby, low luminosity BHs, such as
those in globular clusters.
I also examine the fossil evidence of gas accretion contained in the
kinematics of the stars in the nucleus. The higher-order moments ($h3$
and $h4$) of the line-of-sight velocity distribution show patterns
that are remarkably similar to those seen on larger scales in
elliptical galaxies and in gas-rich merger simulations.  The
kinematics suggests the presence of two components in the M32 nucleus,
a dominant disk overlying a pressure supported component. I discuss
possible formation scenarios for the M32 nucleus in the context of the
kinematic data as well as previous stellar population studies.  The
kinematic measurements presented here are the highest quality
available for the nucleus of M32, and may be useful for any future
dynamical models of this benchmark system.

\end{abstract}

\keywords{galaxies:nuclei -- galaxies:active -- galaxies: kinematics and dynamics -- galaxies:formation -- galaxies: individual (M32) }

\section{Introduction}

The M32 nucleus hosts the smallest of the three supermassive black
holes known in the Local Group.  Like its counterpart in M31 and the
Milky Way, it is currently accreting at very low levels, detectable
only due to its proximity.  Surrounding the black hole (BH) in M32 is
a rotating disk of stars that is the densest stellar system in the
local universe.  In this paper I use high resolution spectroscopic
observations to examine the current accretion of the M32 black hole
and study the formation of the stellar nucleus from its present-day
kinematics.

The kinematic signature of a supermassive BH at the center of M32 was
first recognized by \citet{tonry84}, and current estimates suggest a
BH mass of 2.5$\times$10$^6$~M$_\odot$
\citep{verolme02,vandenbosch10}.  Accretion onto this BH has been
searched for at many wavelengths, but detected only in the X-rays
\citep{ho03}.  The X-ray luminosity from 2-10 keV is
9.4$\times$10$^{35}$ erg$\,$s$^{-1}$, suggesting an accretion rate nine
orders of magnitude below the Eddington limit.  Like M31 and the Milky
Way, the M32 BH presents a rare opportunity to study BH accretion
processes at very low levels.  In \S3, 
I report detection of hot dust emission from near the black hole at
NIR wavelengths with luminosity exceeding the X-ray luminosity by two
orders of magnitude.

In the second half of the paper (\S4), I use the kinematics of the
nuclear region to constrain the formation history of its stars.  The
formation history of stars in the immediate vicinity of the BH is
likely to be connected in some way to the formation of the BH itself
\citep[e.g.][]{hopkins10a}.  The nucleus of M32 contains a rotating
disk of stars that reaches a density at $r < 0.1$~pc of
$>$10$^7$~M$_\odot$/pc$^{-3}$, making it the densest stellar system
known \citep{walker62,lauer98}.  Morphologically, the nuclear
component appears to be distinct from the rest of the galaxy, with the
surface brightness profile showing evidence for an additional stellar
component at radii $<$5\asec~(20~pc) \citep{kormendy99b,graham09}.
This component has an effective radius of 1.5\asec, and an $I$ band
luminosity of $\sim$3$\times$10$^7$~M$_\odot$ \citep{graham09},
$\sim$10\% of the total galaxy luminosity \citep[see
  also][]{kormendy09}.  The mass and size typical of this component is
typical of nuclear star clusters commonly seen at the centers of lower
luminosity early-type galaxies, but it contains a much larger fraction
of the total galaxy luminosity than usual \citep[e.g.][]{cote06}.

Despite the evidence for a morphologically distinct component in the
central 5\asec, no corresponding separation is seen in the stellar
populations or kinematics.  The rotation velocity decreases slowly
with radius \citep{dressler88} but there is no break suggesting a
separation between the nuclear disk and the rest of the galaxy.  The
stellar populations of the nucleus are younger, than the galaxy as a
whole, with an average age of $\sim$4 Gyr in the nucleus rising to
$\sim$8 Gyr at larger radii \citep{worthey04,rose05,coelho09}.
However, again there is no evidence for a break in the stellar
population properties as a function of radius.  In \S4 I show that
there is a kinematic signature in the nucleus that is remarkably
similar to merger simulations done at larger scales.  This signature
provides strong evidence for a two component system in the M32
nucleus, with a dominant disk component embedded in a
pressure-supported component.  I outline a scenario, in which
the nuclear disk formed gradually from the stellar winds of stars in
the bulge of M32, that may provide a good explanation of the observed
kinematics, stellar populations and abundance gradients seen in the
nucleus.


\section{Data}

Integral field spectra of the central 3\asec$\times$3\asec~of M32 were
obtained with Gemini NIFS on Oct 23, 2005.  A total of 11$\times$600s
$K$ band exposures were taken, with 6 on source and 5 offset to a
blank sky position.  
The nucleus was used as a natural guide star to obtain an adaptive
optics correction with the ALTAIR system.  Two previous papers have
utilized this same dataset \citep{davidge08,davidge10} but were
restricted to the study of stellar populations in the nucleus.

Data were obtained from the Gemini Science Archive and reduced
following the same process outlined in \citet{seth08b} and
\citet{seth10}.  An A0V star, HIP~116449, was used as a telluric
calibrator.  This star was also used for photometric and
spectrophotometric calibration, both of which have errors of
$\sim$10\%.  Wavelength calibration utilized both an arc lamp image
and the sky lines in the spectrum, and has an absolute error of
$\sim$2 km$\,$s$^{-1}$.  The median spectral resolution was
4.2\AA~(57~km$\,$s$^{-1}$) FWHM.  The signal-to-noise ratio of the
spectrum is very high, ranging from 280 per
0$\farcs$05\asec$\times$0$\farcs$05 pixel in the center to 60 in the
corner of the field ($r = 2\farcs1$).

To determine the point spread function (PSF) of the observations I
followed a two-step process similar to that described in
\citet{seth10}.  I assumed a PSF described by an inner Gaussian core
$+$ outer Moffat halo (with the Moffat function described as
$\Sigma(r) = \Sigma_0/[(1+(r/r_d)^2]^{4.765}$).  I used the
observation of the telluric calibrator to measure the outer halo of
the PSF and found a best fit $r_d=0\farcs83$~containing 55\% of the
light.  I then fixed this outer halo, and fit the inner part of the
galaxy PSF by convolving an HST image of M32 to fit our NIFS image
(after making a minor correction for the dust emission, see \S3 for
details).  For the HST image, I used an F1042M WFPC2 image where the
nucleus was located on the planetary camera.  I found a very good fit
to the NIFS image by convolving this HST image using an inner Gaussian
with FWHM of 0$\farcs$25 and containing 45\% of the light (along with
the outer Moffat with $r_d=0\farcs83$); typical fit residuals were
4\%.  The compact dust emission discussed in the next section has a
FWHM of 0$\farcs$25, an independent confirmation of the PSF core
width.  Measurements of the CO bandhead strength used in \S3 and
plotted in Fig.~\ref{contcofig} were made in each spaxel after
dividing by the telluric star spectrum using the $^{12}$CO$\,(2,0)$
index defined by \citet{kleinmann86}.

I derived kinematics from the NIFS data from the CO bandhead between
2.28$\mu$m and 2.395$\mu$m using the penalized-pixel fitting (PPXF)
method of \citet{cappellari04} to determine the line-of-sight velocity
distribution (LOSVD).  The program parameterizes the LOSVD using the
radial velocity ($V$), dispersion ($\sigma$), skewness ($h3$) and
kurtosis ($h4$).  High resolution templates from \citet{wallace96}
were used and convolved to the spectral resolution of each pixel as
determined from sky lines \citep[see][for more details]{seth10}.

The measurements presented here are the highest quality kinematic data
available that resolves the sphere-of-influence of the BH.  The very
high S/N results in errors on the LOSVD that are much smaller than for
the HST/STIS measurements presented by \citet{joseph01}. The spatial
resolution of the NIFS observations is somewhat lower than with STIS,
but NIFS provides integral field coverage that can help significantly
in constraining dynamical models \citep[e.g.][]{shapiro06,neumayer07,nowak08}.
Our kinematic measurements are consistent with those of
\citet{joseph01} except in the central few tenths of an arcsecond,
where the higher resolution of STIS results in a steeper velocity
gradient and a central dispersion higher by $\sim$10\%.  As M32 is a
benchmark for studies of BH mass measurements and dynamics codes
\citep{vandermarel98,verolme02,kormendy04b,vandenbosch10}, I make the
full kinematic data set available in Table~1 for use in future
studies.

\begin{deluxetable*}{llcccc}
\tabletypesize{\scriptsize}
\tablecaption{Gemini/NIFS Kinematic Data of the M32 Nucleus}
\tablehead{ 
\colhead{$\Delta\alpha$ [\asec]} & 
\colhead{$\Delta\delta$ [\asec]} & 
\colhead{$V_r$ [km$\,$s$^{-1}$]} & 
\colhead{$\sigma$ [km$\,$s$^{-1}$]} & 
\colhead{$h3$}  & 
\colhead{$h4$}
}
\startdata
  0.014 & -0.015 & -206.14$\pm$1.08 & 117.94$\pm$1.23 &  0.018$\pm$0.007 &  0.038$\pm$0.007 \\
  0.014 &  0.035 & -215.93$\pm$1.09 & 114.63$\pm$1.17 &  0.022$\pm$0.007 &  0.053$\pm$0.007 \\
 -0.036 & -0.015 & -208.52$\pm$1.10 & 115.40$\pm$0.87 &  0.022$\pm$0.005 &  0.043$\pm$0.007 \\
\multicolumn{6}{c}{(\em See Online Journal Article for Full Data Table)}
\enddata
\end{deluxetable*}

\section{Hot Dust: the Most Luminous Tracer of Current BH Accretion}

The spectral energy distributions (SEDs) of quasar and Seyfert nuclei
in the $K$ band are typically dominated by hot dust with temperatures
of 800-1500~K \citep[e.g.][]{kobayashi93,alonso-herrero96, winge00,
  riffel09}.  High resolution observations with HST NICMOS and
adaptive optics observations have shown that these sources are
typically unresolved, and thus originate quite close to the accreting
BH \citep{quillen01b,riffel10b}. From measurements of the time-lag
between UV and NIR light, the size of this dust-emitting region has
been constrained to be $\ll$1~pc in nearby Seyfert galaxies
\citep{minezaki04,minezaki06}.  The luminosity of the dust emission
correlates strongly with AGN luminosities at other wavelengths
\citep{quillen01b}.  Recently, we found unresolved and possibly
variable hot dust emission in the nucleus of the nearby S0 galaxy
NGC~404 \citep{seth10}, the first detection of compact dust emission
in a low-luminosity (LINER) AGN.

\begin{figure}
\plotone{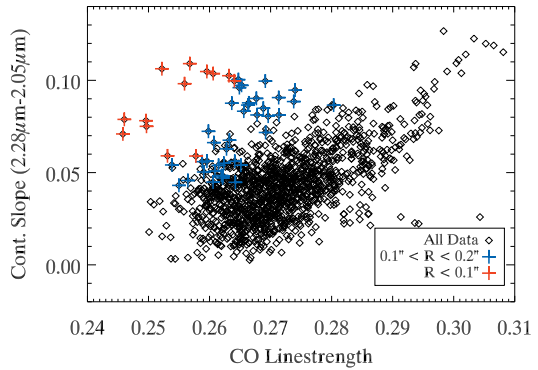}
\caption{The CO line strength plotted against continuum slope.
  Positive values indicate deeper CO lines and redder continuum
  slopes.  Spaxels between 0$\farcs$1 and 0$\farcs$2 of the center are
  shown with blue crosses, and those within 0$\farcs$1 with red
  crosses.  The deviation towards redder colors and weaker CO lines is
  due to emission from hot dust.  The slope is measured from the flux
  ratioed spectra, thus a value of zero indicates the same spectral
  slope as an A0 star.  Errors on the CO depths are $\lesssim$0.005
  and on the continuum slopes, $\lesssim$0.0002.}
\label{contcofig}
\end{figure}

The presence of dust emission in the nucleus of M32 is inferred due to
the presence of a redder continuum and weaker CO lines in the central
0$\farcs$2.  This is shown in Fig.~\ref{contcofig}.  Typically, the
color of the continuum correlates with the depth of the CO line as
cooler redder stars have deeper CO lines.  However, within the central
0$\farcs$2, (and especially the central 0$\farcs$1 shown with red
crosses in Fig.~\ref{contcofig}), the continuum gets redder while the
CO lines become weaker indicating the presence of hot dust emission.
To determine the contribution from this hot dust component I fit the
M32 nuclear spectrum to a combination of a blackbody model and an
annular spectrum from radii of 0$\farcs$5-1\asec~\citep[as in
  Fig.~4 of][]{seth10}.  The best fitting temperature for the dust at
the center ($r < 0\farcs1$) is 920K.  The residuals of the fit are
1.1\%, a factor of two improvement over a fit with no dust.

I fit the dust contribution to each spaxel in the data cube fixing the
dust temperature to 920K.  Dust emission is only found at
$r\lesssim$0$\farcs$2 and reaches a maximum contribution of $\sim$5\%
of the total light.  The dust emission is unresolved, with a FWHM of
0$\farcs$25 matching the NIFS PSF derived from HST observation.  The
emitting region is therefore constrained to be $<$0.9~pc.  

The total flux of the hot dust emission from 2.0-2.43~$\mu$m is
1.7$\times$10$^{-13}$~erg$\,$s$^{-1}\,$cm$^{-2}$ corresponding to a
luminosity in the $K$ band of 1.2$\times$10$^{37}$~erg$\,$s$^{-1}$.
Integrating over the full Planck function, the total luminosity is
1.8$\times$10$^{38}$~erg$\,$s$^{-1}$.  The corresponding $K$ band
magnitude of the emission is $K$=15.0 ($M_K=-9.4$).  The primary
uncertainty in these flux and luminosity measurements is the
temperature of the dust, which is not well constrained by our fits.
Taking a reasonable range of temperatures from 800-1300K, the flux
ranges from 1.5-2.5$\times$10$^{-13}$~erg$\,$s$^{-1}\,$cm$^{-2}$, the
$K$ band luminosity from 1.0-1.7$\times$10$^{37}$~erg$\,$s$^{-1}$ and
the total luminosity from 1.3-2.6$\times$10$^{38}$~erg$\,$s$^{-1}$.

The weaker CO lines and redder continuum cannot result just from
changes in the stellar population at the center of M32.  There is
little indication of any change in the optical and UV colors near the
center of M32, although slightly bluer colors are seen in the central
pixels of a WFPC2 F555W-F814W color image by \citet{lauer98}.  The
Galactic center also shows a decrease in CO line depth within the
central parsec \citep{sellgren90} which likely results from the
decrease in the fraction of light coming from late-type stars in
the central parsec \citep{genzel96,do09,bartko10}.  This decrease may
be caused by stellar collisions preferentially destroying the larger
radius RGB and AGB stars \citep[e.g.][]{dale09}.  The M32 stellar
density is sufficient that such collisions may occur there as well
\citep{lauer98}.  However, this mechanism cannot by itself explain the
effect shown in Fig.~\ref{contcofig}, as the removal of larger radius
RGB and AGB stars would make $K$ band continuum bluer, not redder.

Emission from hot dust could also result from heating by the stellar
component.  M32 has a population of post-AGB and extreme Horizontal
Branch stars \citep{brown98,brown00}, but these are much less
compactly distributed than the dust emission.  Furthermore, the total
UV luminosity at 1600\AA~is insufficient to heat the dust \citep[$\nu
  L_\nu \sim 2\times10^{37}$~erg$\,$s$^{-1}$; ][]{cole98}, and the
UV-to-optical colors do not decrease at the center of M32
\citep{lauer98}.  Therefore hot stars do not appear to be responsible
for the dust emission.  The interacting stellar winds resulting from
the high stellar density and velocity dispersion could also provide
energy to heat the dust.  The conditions at the center of M32 are
fairly similar to those in the MW center where no comparable hot dust
component is seen.  However, there is a significant warm dust
component in the Milky Way center with T$\sim$400~K and luminosity of order
10$^{38}$~erg$\,$s$^{-1}$ within the central parsec \citep{zylka95,
  stolovy96,mezger96}.  I also note that there are no signatures of
compact hot dust emission in NIFS observations of local group galaxies
M33 and NGC~205 which also have high stellar densities (but lower
dispersions).  Based on the lack of hot dust emission in these
systems, I argue that the hot dust in the nucleus of M32 is likely
powered by BH accretion.

\begin{figure*}
\plotone{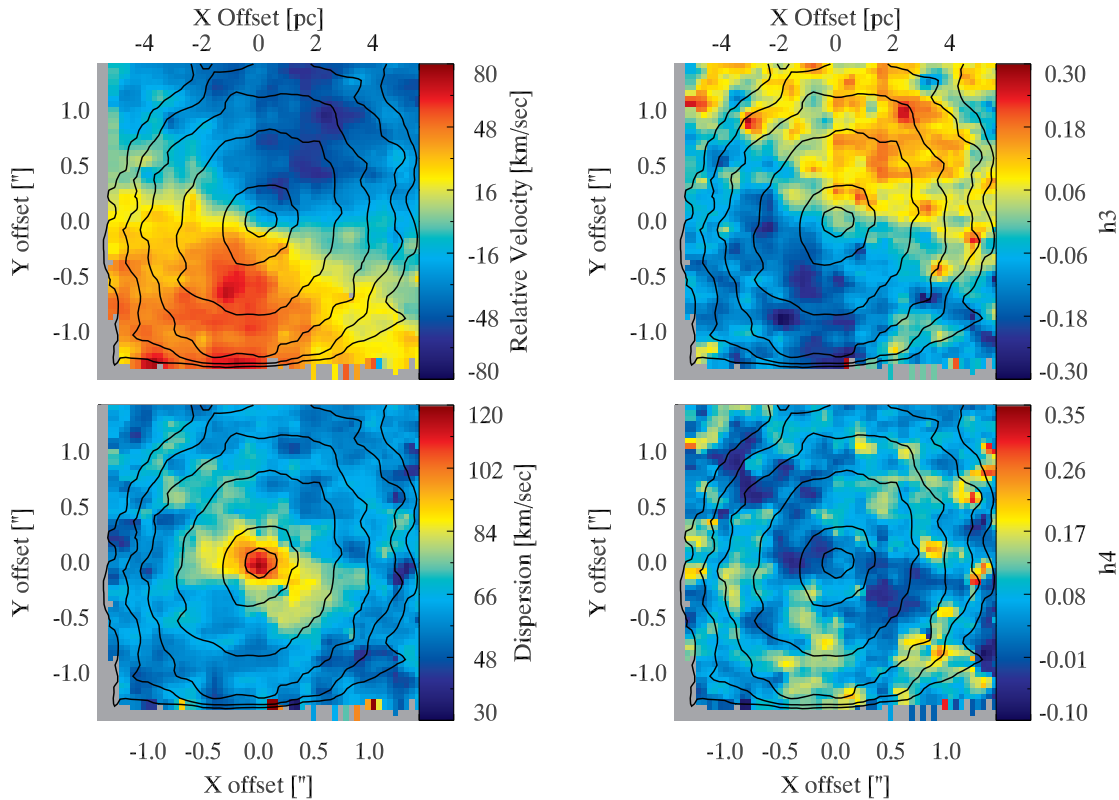}
\caption{The line-of-sight velocity distribution of M32 derived from
  the Gemini/NIFS data at the CO bandhead (2.3$\mu$m).  The radial
  velocity is shown in the top left panel after removal of the
  systemic velocity of -200~km$\,$s$^{-1}$.  The bottom left panel
  shows the dispersion, the top right the skewness or $h3$ component,
  and the bottom right the kurtosis or $h4$ component of the LOSVD.
  These data are the highest S/N data available of the M32 nucleus
  that resolve the BH sphere of influence.}
\label{allkinfig}
\end{figure*}

The compact hot dust emission appears to be the most luminous BH
accretion signature observed in the M32 nucleus. \citet{ho03} analyzed
a variety of radio and X-ray data and detect the nucleus only in
X-rays, with a 2-10~keV luminosity of
9.4$\times$10$^{35}$~erg$\,$s$^{-1}$.  Based on this X-ray luminosity
and the expected supply of gas available for accretion \citet{ho03}
found that the radiative or BH accretion efficiency must be very small
($\lesssim10^{-3}$).  Such low efficiencies are found in radiatively
inefficient accretion models \cite[e.g.][]{quataert01,ho09}.  The NIR
emission discussed here would allow for a somewhat higher accretion
efficiency, but still likely lower than the 10\% efficiency of a
canonical accretion disk.  This suggests that strong NIR dust emission
may be found in lower accretion states as well as the higher accretion
states where they have traditionally been observed.  The dust emission
in Seyferts and quasars is thought to come from a radiatively heated
sublimation zone in the accretion disk or at the inner edge of the
torus.
However, the excitation mechanism for the hot dust emission in M32 is
unclear.  The NIR luminosity derived here is two orders of magnitude
brighter than the observed X-ray luminosity found by \citet{ho03}.  In
studies of other low luminosity AGN, the NIR to X-ray luminosity
ratios are typically unity \citep{ho99,ho08,eracleous10}.  These
systems are somewhat higher luminosity systems than M32, but all are
expected to be radiatively inefficient accretors with similar spectral
energy distributions \citep{ho09}.  Furthermore, the observed X-ray
luminosity and UV upper limits \citep{cole98,ho03} are insufficient to
supply the radiative energy needed to heat the dust.  Both these
problems could be solved if the BH accretion in M32 is significantly
time variable, with our observations taken at a time when there was
higher flux at all wavelengths.  Alternatively, the heating of the hot
dust could result from mechanical feedback from the BH
\citep[e.g.][]{clenet05} or possibly stellar winds as discussed above.



Based on the limited examples of M32 and NGC~404~\citep{seth10}, it
appears NIR spectra at high spatial resolution may be an effective way
to search for accretion in the lowest luminosity AGN.  In both these
cases, the NIR emission is the most luminous activity tracer, more
luminous than the detected X-ray emission.  Furthermore, the low
luminosity of the X-ray emission ($\lesssim10^{38}$~erg$\,$s$^{-1}$) can be
confused with emission from X-ray binaries
\citep[e.g.][]{desroches09,gallo10} and thus the NIR emission may
represent the best indicator of BH accretion if it can be spectrally
disentangled from the stellar light.  One very interesting application
of this technique would be to search for hot dust emission from
accreting BHs in G1 and other globular clusters with putative central
BHs \citep[e.g.][]{ulvestad07,noyola10}.


\begin{figure*}
\plotone{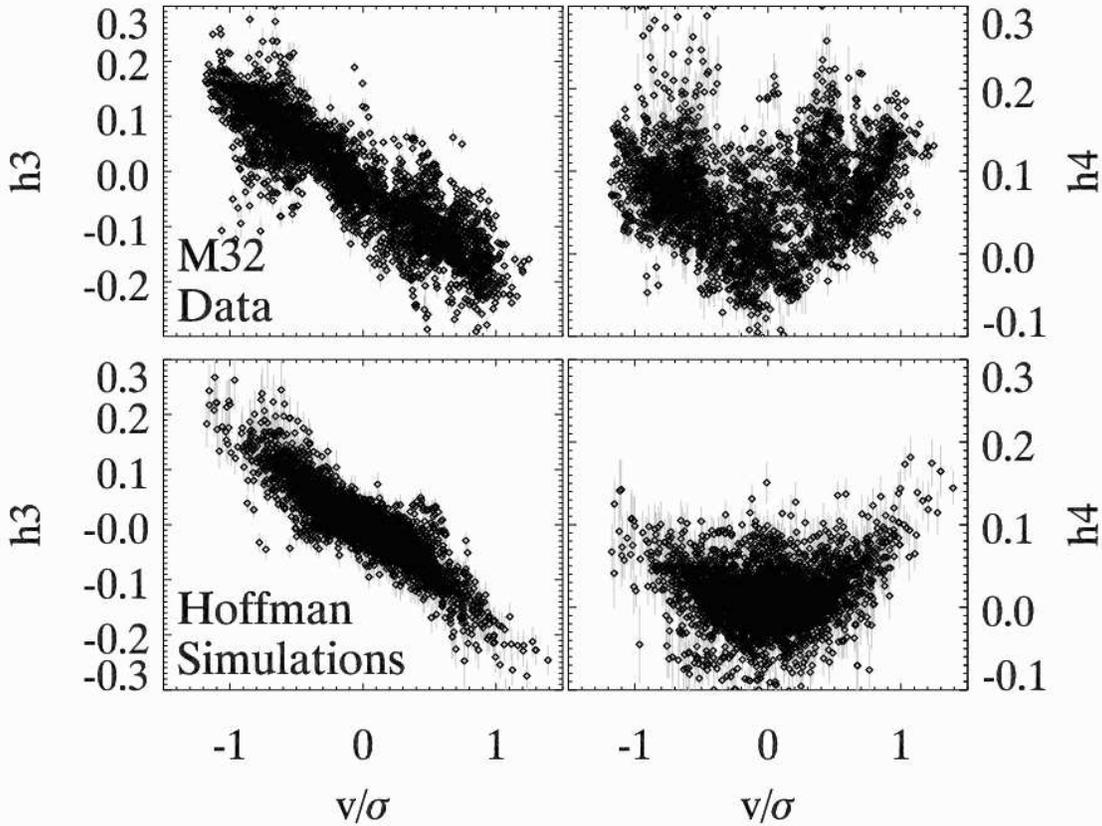}
\caption{Distribution of the higher order LOSVD moments ($h3$ and
  $h4$) as a function of the rotation velocity divided by the
  dispersion $V/\sigma$ at $r < 1.5$\asec.  The top panels display the
  Gemini/NIFS kinematics of M32, while the bottom panels display data
  from gas-rich mergers at similar orientations by \citet{hoffman09}.
  In the top panels, each spaxel in our IFU observations is plotted as
  a single data point (with a light gray error bar), while the
  simulation data has been analyzed in a similar way to mimic IFU data.\\
}
\label{momentsfig}
\end{figure*}

\section{A Kinematic Signature of Past Gas Accretion}

In this section I discuss kinematic measurements from the CO bandhead
of the NIFS data.  The higher-order moments of the LOSVD show a very
clear signature similar to that typically seen on larger scales in
kinematic measurements of the centers of rotating, disky elliptical
galaxies \citep[e.g.][]{bender94,krajnovic08}.  The kinematic
signature seen in these galaxies has been recently been simulated as
resulting from gas accretion at the centers of galaxies during the end
phases of gas-rich major mergers \citep{hoffman09,hoffman10}.  I
analyze these kinematic results in light of these observations and
simulations.

Maps of the first four moments of the LOSVD derived from our NIFS data
are shown in Fig.~\ref{allkinfig}.  The velocity map shows strong
rotation at a position angle of -25$^\circ$ (E of N) with an amplitude
of $\sim$55~km$\,$s$^{-1}$ beyond a radius of $0\farcs3$
\citep[derived using the kinemetry software of][]{krajnovic06}.  The
dispersion increases to a maximum of 120~km$\,$s$^{-1}$ in the central
$0\farcs5$ due to the influence of the BH and unresolved rotation;
outside of this radius, it drops to values of $\sim$60 km$\,$s$^{-1}$.
The $h3$ distribution is clearly anti-correlated with the velocity,
while a more complex but axially symmetric pattern is visible in the
$h4$ map with negative values along the minor axis and strong positive
values along the major axis.  The effect of individual stars is
clearly seen in the outer parts of the kinematic maps; these result in
lower dispersions, extreme $h3$ values and strongly positive $h4$
values.  The properties of these stars are discussed in
\citet{davidge10}.

The top panels of Figure~\ref{momentsfig} shows the $h3$-$v/\sigma$
and $h4$-$v/\sigma$ distributions of the data.  The $h3$ distribution
is strongly and tightly anti-correlated with $v/\sigma$.  The highest
$h4$ values are found in the most strongly rotating regions of the M32
nucleus.  Both the $h3$ and $h4$ trends are remarkably similar to
those seen on larger kiloparsec scales in observations
\citep{bender94,krajnovic08} and in recent simulations
\citep{hoffman09}.  The $h3$ anti-correlation with $v/\sigma$ is
naturally explained by the presence of some stars centered at the
systemic velocity, while the positive $h4$ values can also result from
the superposition of a cold disk and hot bulge or be a signature of
violent relaxation.  In combination, these trends suggest the presence
of two coincident components in the M32 nucleus, a dominant rotating
disk and a thermalized, relaxed bulge structure.

To facilitate visual comparison between our observations and the
simulations, in the bottom panels of Figure~\ref{momentsfig} I have
plotted models from \citet{hoffman09}.  The plotted models are the
results of 8 gas-rich (40\% gas), equal mass mergers.  The kinematic
data was derived by viewing the central few kiloparsecs of these
simulations from a variety of lines-of-sight with inclinations
consistent with the 70$\pm$5$^\circ$ inclination for the central disk
of M32 found by \citet{verolme02}.  The slope and extent of the
$h3$-$v/\sigma$ distribution is almost identical to that seen in the
simulations.  For $h4$, both the simulations and data have generally
positive $h4$ with a strong similarity in the shapes of the
$h4$-$v/\sigma$ distribution.  The very high $h4$ ($>0.15$) values
seen in the M32 data are the result of individual stars and clumps
super-imposed on the velocity field.

The kinematic data provide evidence for a two component system
including a disk created by gas accretion into the nucleus.  However,
the details of the gas accretion leading to the formation of the
central disk may be somewhat different than in the simulations due to
the large differences in scale between them.  Embedded disk structures
occur in the \citet{hoffman09} simulations only in gas-rich major
mergers and form on short timescales.  However on the smaller scales
of the M32 nucleus, it is possible that lower gas fraction mergers,
gas-rich minor mergers, or even secular processes are still capable of
creating the disky kinematics seen in M32 on parsec scales.  For
instance, in \citet{seth10}, we present evidence that a minor merger
$\sim$1~Gyr ago in NGC~404 appears to be responsible for a burst of
star formation that formed $\sim$5 million solar masses of stars in
the central 10~pc of that S0 galaxy.

Another possibility for formation of nuclear disk is the gradual
accretion of material from stellar winds at the center of the galaxy
as described by \citet{bailey80}.  This would naturally lead to a
gradual trend in stellar age \citep{worthey04,rose05} and the lack of
any break in the rotation curve \citep{dressler88}.  Furthermore, this
scenario may naturally account for the abundance gradient in the
nucleus; \citet{worthey04} and \citet{rose05} find the [Mg/Fe] ratio
is significantly subsolar in the nucleus and increases with increasing
radius.  If the stars in the nucleus were made from the stellar winds
of previous generations of stars, then this might result in a drop in
the [Mg/Fe] \citep[e.g. as in the Milky Way globular cluster
  NGC~2808;][]{carretta09}.  There is sufficient mass in the bulge of
M32 to supply the necessary gas to the nucleus.  However, whether the
gas would actually end up as a rotating structure at the center would
require detailed hydrodynamic simulations.  This scenario could also
be tested using stellar population modeling of resolved stars and
spectra in and around the nucleus if the models are accurate enough to
distinguish between populations of intermediate (2-10~Gyr) age
\citep[e.g.][]{marigo08}.


\subsection{Asymmetries in the M32 Nucleus}

Simulations by \citet{hopkins10b,hopkins10a} follow gas accretion down
to parsec scales in galaxy mergers and find that gas accretion in the
presence of BHs cause the formation of lopsided disks.  These disks
correspond quite well to the eccentric disk seen in M31
\citep[e.g.][]{lauer93}, and \citet{hopkins10a} suggest that this disk
formed at the same time the M31 BH was growing.  Due to the parsec
scale of the expected asymmetries, M32 is one of the few other
galaxies close enough for the asymmetries to be seen.  If the M32
nucleus had a major merger formation scenario similar to that
considered by \citet{hopkins10a}, a lopsided nuclear disk would be
expected.
While M32 clearly does not have the prominent asymmetric structure
similar to that seen at the center of M31, I analyzed HST images to
see if there is any evidence for a lower level asymmetry.

\begin{figure*}
\includegraphics[width=3.4in]{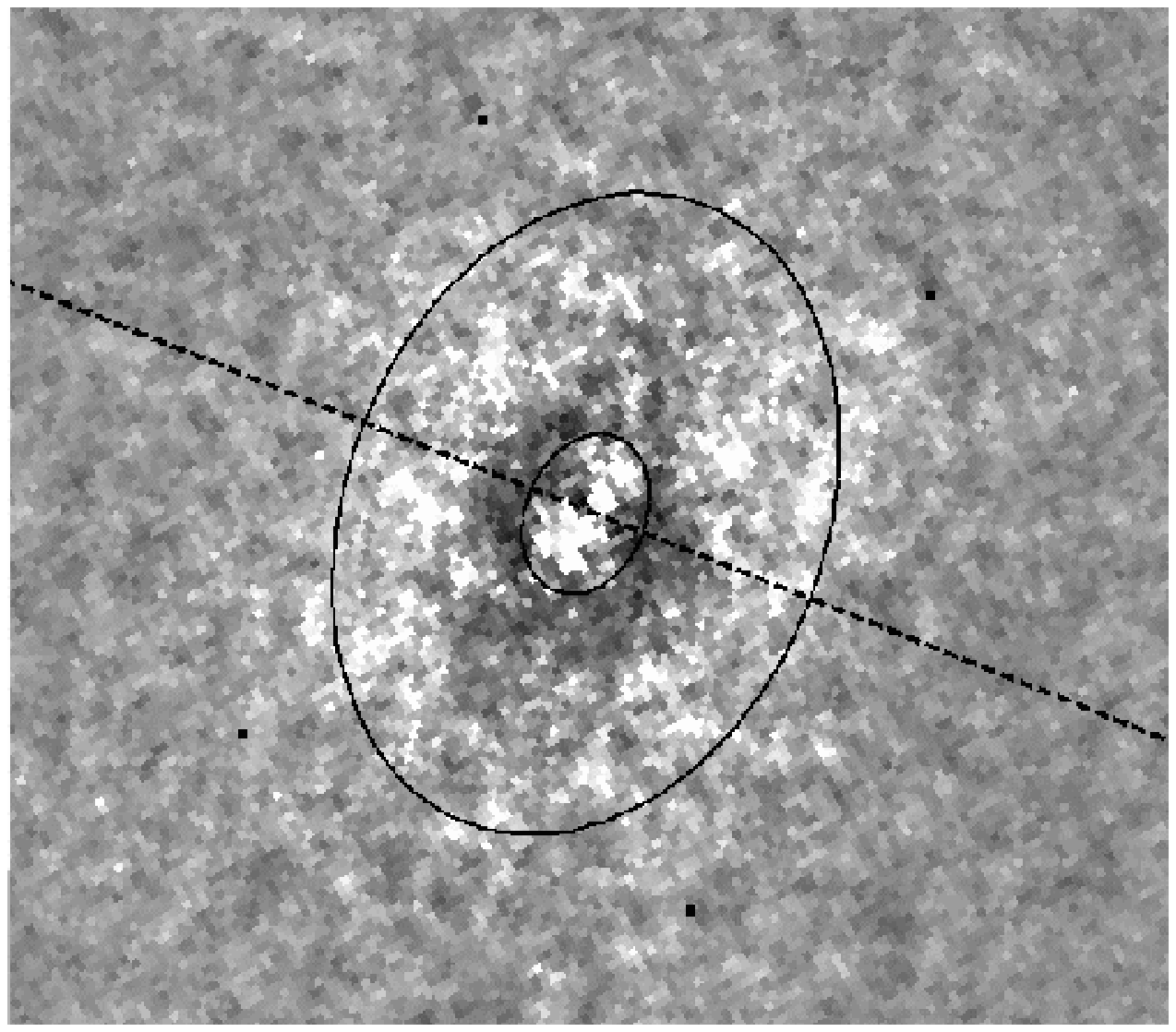}
\includegraphics[width=3.6in]{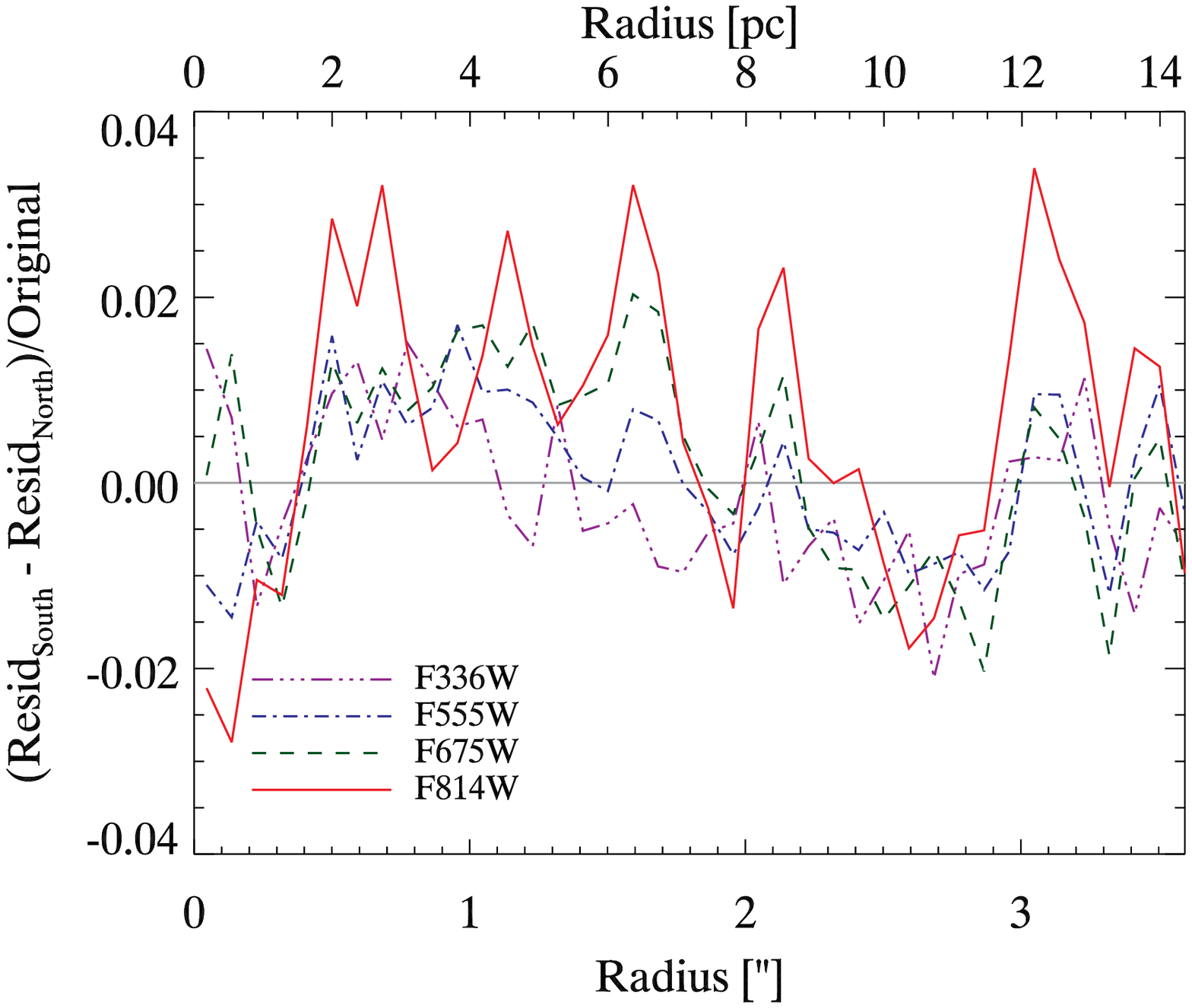}
\caption{Figures showing a small asymmetry in the M32 nucleus.  {\em
    Left --} shows the a model-subtracted F555W image oriented with N
  up and E to the left.  Black solid lines show the annulus with
  major-axes of 0.45\asec~to 1.8\asec, while the dashed line separates
  the northern and southern halves of this annulus.  {\em Right --}
  comparison of the residuals of the model-subtracted images.  The
  Y-axis plots the difference in flux between the southern and
  northern annuli in the model-subtracted images normalized by the
  original image.  A 1-2\% excess in the southern annulus is seen in
  all bands from 0.45$-$1.8\asec (2$-$7~pc).}
\label{residfig}
\end{figure*}

Archival HST WFPC2/PC images were obtained in four bands (F336W,
F555W, F675W, and F814W) from the CADC WFPC2
associations\footnote{http://cadcwww.hia.nrc.ca/hst/wfpc2/wfpc2\_r2.html}.
I then fit axially symmetric model profiles to the data, subtracted
these off and examined the residuals.  Fits to the images in each
filter were performed on the central 20\asec~of each image using
elliptical double S\'ersic profiles.  The outer S\'ersic component was
constrained to have the same indices and effective radii as fits to
the 1-D M32 surface-brightness profile by \citet{graham02} and
\citet{graham09} that include measurements from the same F814W HST
data we use here.  All fits were convolved with an appropriate TinyTim
WFPC2 PSF \citep{krist95}.  The resulting fits accurately described
the light profile at all radii to within 3\%.  The inner S\'ersic
profile fits had effective radii of 1.5-1.6\asec, consistent with that
found by \citet{graham09}, minor-to-major axis ratios of 0.73 and
position angles of $-$20$^\circ$.  After subtraction of these
profiles, some asymmetric residuals were visible (see
Fig.~\ref{residfig}, left panel).  To quantify these, I compared the
total residual flux on one side of the minor axis to the other in
elliptical annuli (Fig.~\ref{residfig}, right panel). At radii between
0$\farcs$45 and 1$\farcs$8 ($\sim$2-7~pc) there is an excess of 1-2\%
of the total flux in the southern half relative to the northern half.
This is seen most clearly in F555W and F675W, and is also present but
less clear in the other two bands due to lower S/N (at F336W) and
increased surface brightness fluctuations (in F814W).  

This small photometric asymmetry is much less lopsided than the
central part of M31 (where the disk eccentricity $\sim$40\%).  The
asymmetry corresponds to a total mass of
$\sim$4$\times$10$^5$~M$_\odot$ derived from the F814W band using the
mass-to-light ratio from \citet{verolme02}.  It is unclear if such
small asymmetries are consistent with the \citet{hopkins10a} model.
The physical situation in M32 is quite different from M31, with a much
smaller BH mass and a higher stellar density.  Simulations that are
more appropriate to M32 may serve to both (1) help understand in
detail how the M32 nucleus formed, and (2) test the feasibility of the
lopsided disk model for driving BH growth in lower mass systems.

\section{Conclusions}

In this paper, I have presented results on the current and past gas
accretion in the M32 nucleus using adaptive optics assisted
Gemini/NIFS data.  The data reveal an unresolved emission component at
the center of M32 that results in the weakening of CO lines and
reddening of the continuum slope in the nucleus.  This emission
appears to originate from hot dust emission close to the BH as seen in
many higher luminosity AGN.  The hot dust emission is unresolved with
a size of $<$0.9~pc and has a luminosity of $\sim$2$\times10^{38}$
erg$\,$s$^{-1}$.  This luminosity is more than two orders of
magnitude more luminous than the X-ray emission detected from the
central BH by \citet{ho03}, the only other bandpass in which emission
due to accretion has been detected in this galaxy.  Combined with
similar emission in the nearby low luminosity LINER NGC~404, this suggests
that high resolution NIR spectra may be an efficient way of locating
low level activity from BHs in nearby galaxies and globular clusters.

I also present the highest quality kinematics of the M32 nucleus
currently available, potentially useful for any future dynamical
studies of this benchmark system.  These measurements cover the
central 3\asec$\times$3\asec~with a resolution of 0$\farcs$25.  The
very high signal-to-noise of these measurements allows accurate
determination of the higher order moments which reveal a remarkable
kinematic signature similar to those seen on larger scales in
elliptical galaxies and merger simulations.  The signature manifests
as an anti-correlation of $h3$ with $v/\sigma$ and strongly positive
$h4$ especially in areas of strong rotation.  These kinematics suggest
the presence of a dominant disk component embedded in a pressure
supported system.  I propose a possible formation scenario in which
the M32 nucleus forms from stellar winds of the M32 bulge.  This
scenario qualitatively explains the kinematic, stellar population and
abundance properties of the nucleus.  I also briefly consider whether
M32 contains any asymmetries in its nucleus, as might be expected
based on the \citet{hopkins10a} simulations of major mergers.  A small
$\sim$2\% photometric asymmetry is found at radii between 2 and 7~pc.

Acknowledgments: The author would like to thank the referee, Luis Ho,
for improving this paper.  He also thanks Loren Hoffman for sharing
her simulation data, and to thank her, Michele Cappellari, Davor
Krajnovi\'c, Jay Strader, Nelson Caldwell, Knut Olsen, Hagai Perets,
and Margaret Geller for helpful discussions.  The author is supported
by a fellowship from the Smithsonian Institute.

{\it Facilities:} \facility{Gemini:Gillett (NIFS/ALTAIR)}, \facility{HST (WFPC2)}\\


\end{document}